\documentclass[aps,jcp,twocolumn,showpacs,groupedaddress]{revtex4}
\usepackage{graphicx,amssymb,amsmath}
\usepackage{epstopdf}
\begin{document}
\title{Simple crystallizable bead-spring polymer model}
\author{Robert S. Hoy}
\email{rshoy@usf.edu}
\affiliation{Department of Physics, University of South Florida, Tampa, FL 33620}
\author{Nikos Ch. Karayiannis}
\affiliation{Institute for Optoelectronics and Microsystems (ISOM) and ETSII, Universidad Polit{\'e}cnica de Madrid, Madrid, Spain}
\pacs{64.70.km,64.60.Cn,64.70.Dg,83.80.Ab} 
\date{\today}
\begin{abstract}
We develop a simple coarse-grained bead-spring polymer model exhibiting competing crystallization and glass transitions.  For quench rates slower than the critical nucleation rate $|\dot{T}|_{crit}$, systems exhibit a first-order crystallization transition below a critical temperature $T=T_{cryst}$.  Such systems form close-packed crystallites of FCC and/or HCP order, separated by domain walls, twin defects, and an amorphous interphase.  The size of amorphous regions grows continuously as the quench rate $|\dot{T}|$ increases, producing nearly amorphous structure for  $|\dot{T}|>|\dot{T}|_{crit}$.  Our model exhibits many features observed in recent studies of crystallization of athermal polymer packings, but also critical differences arising from the softness of the pair interactions and the thermal nature of the phase transition. The model is considerably more computationally efficient than other recent crystallizable coarse-grained polymer models; while it sacrifices some features of real semicrystalline polymers (such as lamellar structure and chain disentanglement), we anticipate that it will serve as a useful model for studying generic features related to semicrystalline order in polymer solids.
\end{abstract}
\maketitle

\section{Introduction}
\label{sec:intro}

A coarse-grained polymer model should include the minimal set of features necessary to capture the physical phenomena of interest while remaining maximally computationally expedient.  For example, the flexible Kremer-Grest (KG) bead-spring model\cite{kremer90} is a minimal model including only chain connectivity, excluded volume and van der Waals attractions.  Despite this simplicity, it is able to capture the behavior of real polymers to an extraordinary degree, exhibiting features ranging from Rouse and entangled dynamics (i.e. reptation\cite{doi86}) in its molten state\cite{kremer90,putz00} to dynamical heterogeneity \ in its glass transition regime\cite{gebremichael01} to aging, rejuvenation, and strain hardening in its amorphous glassy state.\cite{rottler05,hoy06,warren07}

One limitation of the standard KG model is that it possesses an inherent length-scale competition; the equilibrium length $\ell_0$ of covalent bonds is significantly different from the equilibrium separation $r_0$ for nonbonded monomers.
This competition prevents formation of the semicrystalline order possessed by most real polymers.
United atom models\cite{ryckaert78,toxvaerd90} exhibit crystallization\cite{liu98} as well as glass formation,\cite{pant93} and include the angular and dihedral interactions required to map to specific polymer chemistries, but are computationally expensive.
In the opposite limit, the simplest models treat polymers as freely-jointed chains of tangent hard spheres with $\ell_0=r_0$ and have recently illustrated the competition between athermal glass formation (jamming) and crystallization.\cite{karayiannis08,karayiannis09b,karayiannis09,karayiannis10,karayiannis13}
The limitation of these latter, highly-idealized models, of course, is that they are athermal, while $k_B T$ is a critical parameter that profoundly affects polymer properties.
It is desirable, therefore, to develop simple models which possess both the soft excluded volume and van der Waals attractions necessary to capture thermal behavior (in particular, exhibiting a glass transition) \textit{and} a local chain structure amenable to crystallization, e.g.\ $\ell_0 = r_0$.

In this paper we develop and describe the basic properties of such a polymer model.
We will show that rapidly quenched systems remain largely amorphous down to $T=0$ while slowly quenched systems display a degree of crystalline order that increases with decreasing quench rate $|\dot{T}|$.
Consistent with results for athermal polymer packings\cite{karayiannis09,karayiannis10,karayiannis13}, our model forms close-packed crystallites of face centered cubic (FCC), hexagonal close packed (HCP), or mixed FCC/HCP order with varying degrees of stacking faults and five-fold-symmetric defects.
While real semicrystalline polymers typically do not form close-packed crystals, we will show that our model captures generic features of polymer crystallization.

\section{Model and Methods}
\label{sec:modelmeth}

Each polymer chain contains $N=50$ coarse-grained beads, while each bead corresponds to 2-5 monomers.\cite{kremer90}
All beads have mass $m$ and interact via the truncated and shifted Lennard-Jones potential
$U_{\rm LJ}(r) = 4u_{0}[(a/r)^{12} - (a/r)^{6} - (a/r_{c})^{12} + (a/r_c)^{6}]$,
where $a$ is monomer diameter, $r_{c}$ is the potential cutoff radius, and $U_{LJ}(r) = 0$ for $r > r_{c}$.
The unit of time is $\tau = \sqrt{ma^{2}/u_{0}}$ and maps to time scales in the 10-100ps range; \cite{kremer90} we employ a timestep $\delta t = \tau/300$.
We set $r_0 = \ell_0 = a$ by choosing the pair interactions $U_{\rm LJ} = 4u_0[(\sigma/r)^{12} - (\sigma/r)^{6} - (\sigma/r_c)^{12} +(\sigma/r_c)^{6}]$, with $\sigma=2^{-1/6}a$ and $r_c=2^{7/6}a$, and using a stiff harmonic bond potential of form $U_{bond}(\ell) = (k_b/2)(\ell-a)^2$, with $k_b = 600\epsilon$.
The energetic barrier to chain crossing is $k_b(\sqrt{2}-1)^2\simeq100u_0$, i.e.\ $\gtrsim 100k_BT$ for the systems considered here.
Systems consist of $N_{ch}=500$ chains and periodic boundary conditions are applied in all three directions.
Intial melt states are generated with a monomer number density $\rho=1.0a^{-3}$ (packing fraction $\phi = \pi\rho/6$).
After thorough equilibration at $k_BT=1.2u_0$, systems are quenched to zero temperature at various rates $|\dot{T}|$ while maintaining zero hydrostatic pressure using a Nose-Hoover barostat. Simulations are performed using LAMMPS.\cite{plimpton95}
Throughout the rest of the paper, we will express temperatures in units of $k_BT/u_0$, quench rates in units of $\tau^{-1}$, distances in units of $a$ and densities in units of $a^{-3}$.

During the quenches we monitor several quantities including the potential energy per monomer $U$, pair correlation function $g(r)$, packing fraction $\phi$, and metrics of local structure including the Characteristic Crystallographic Element (CCE) norm.\cite{cce09,karayiannis09,karayiannis10,karayiannis13}  The later is a highly discriminating descriptor which quantifies the orientational and radial similarity of a local environment to a given ordered structure in atomic and particulate systems. The CCE norm is built around the defining set of crystallographic elements and the subset of distinct elements of the corresponding point symmetry group that uniquely characterize the reference crystal structure. For example, the FCC crystal symmetry is mapped onto a set of four three-fold axes (roto-inversions of $2\pi/3$), while the HCP is mapped onto a single six-fold symmetry axis (roto-inversion of $\pi/3$). A scan in the azimuthal and polar angles identifies the set of axes that minimize the CCE norm of a reference site (atom or particle) with respect to a crystal structure $\emph{X}$.  Details on the underlying mathematical formula and the algorithmic implementation can be found in Ref. \cite{cce09}. Once the CCE norm ($\epsilon_{i}^{X}$) is calculated for each site $\emph{i}$ a corresponding order parameter $s^{X}$ can be calculated which is practically equal to the fraction of sites with CCE norms below a pre-set threshold value ($\epsilon_{i}^{X}\le\epsilon^{\rm thres}$).
Results from the CCE-norm-based analysis with respect to FCC, HCP, and fivefold symmetries are presented below.

\section{Results}
\label{sec:results}

\begin{figure}[htbp]
\includegraphics[width=3in]{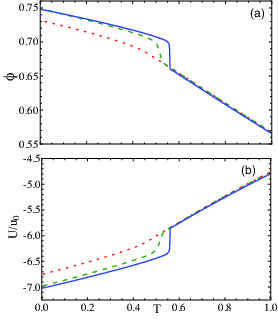}
\caption{Volumetric and energetic measures of the crystallization transition.  Solid blue, dotted green, and dashed red lines show data for $|\dot{T}|=10^{-6}$, $|\dot{T}|=10^{-5}$, and $|\dot{T}|=10^{-4}$, respectively.    Panel (a) illustrates the packing fraction $\phi$ and panel (b) illustrates the potential energy per monomer $U$.  For the slowest quench rates, both data sets indicate $T_{cryst}\simeq0.56$.}
\label{fig:phiandU}
\end{figure}

Figure \ref{fig:phiandU} illustrates the evolution of packing fraction $\phi(T)$ and potential energy $U(T)$ at various quench rates.
For slow quench rates, the data show clear signatures of a crystallization transition at $T_{cryst}\simeq 0.56$; $\phi$ ($U$) exhibit upward (downward) jumps that indicate an increasingly first-order-like transition as $|\dot{T}|$ decreases.
As $T$ approaches zero, $\phi$ approaches the maximal value for close packed crystals, $\phi_{\rm cp}=\pi/\sqrt{18}=.7405$.
Indeed, at the slowest $|\dot{T}|$, $\phi$ exceeds $\phi_{\rm cp}$; this is attributable to the softness  and long-range attractive tail of $U_{\rm LJ}$.
In contrast to recent work on athermal systems\cite{karayiannis08,karayiannis09b}, our model does not ``jam'' at random close packing  ($\phi_{RCP}=.636$;\cite{torquato00}) like atomic Lennard Jones systems\cite{broughton82}, it is an excellent crystal-former.
For the fastest quench rate, $\phi$ and $U$ show no apparent crystallization transition, and a weak glass transition, as indicated by a smooth bend in $U$ and $\phi$, is observed at $k_BT = T_g \simeq 0.45u_0$.  Thus, as in real semicrystalline polymers, according to the proposed model $T_{cryst} > T_g$.

\begin{figure}[htbp]
\includegraphics[width=3in]{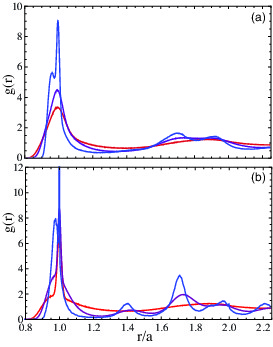}
\caption{Pair correlation functions at characteristic temperatures and different quench rates.  Red, purple and blue lines  respectively show data for $k_B T/u_0 = 1.0,\ 0.5\ \rm{and}\ 0$.Panel (a) shows data for $|\dot{T}|=10^{-4}$ while panel (b) shows data for $|\dot{T}|=10^{-6}$}
\label{fig:gofr}
\end{figure}

Figure \ref{fig:gofr} shows the evolution of the pair correlation function $g(r)$ with $T$ at the slowest and fastest $|\dot{T}|$.
Results are shown for temperatures well above the melting point, slightly below $T_{cryst}$, and zero.  Above the melting point, systems have amorphous (melt-like) structure as expected.
For slow quenches, just below $T_{cryst}$, peaks in the correlation function form corresponding to the appearance and growth of close packed order.
At zero temperature, clear peaks at the characteristic second and third nearest neighbor distances for close-packed crystals, $r_{2n}=\sqrt{2}$ and $r_{3n}=\sqrt{3}$ have developed\cite{footdoubpeak}; the system also retains some amorphous character as indicated by the large width of these peaks.
In sharp contrast, for fast quenches $|\dot{T}|=10^{-4}$, systems at the same temperatures remain predominantly amorphous; $g(r)$ maintains liquid-like structure down to $T=0$.

\begin{figure}[htbp]
\includegraphics[width=3in]{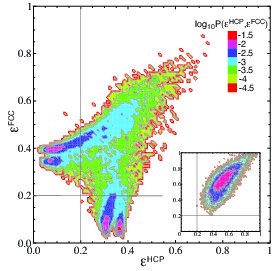}
\caption{Parity plot of the $\epsilon^{\rm FCC}$ versus $\epsilon^{\rm HCP}$ CCE-based norms over all monomers for $|\dot{T}|=10^{-6}$ at $T=0$ and (inset) $T=1.0$. Colors indicate the (log-scale) site ordering probability density $P(\epsilon^{\rm HCP},\epsilon^{\rm FCC})$.  Horizontal and vertical gray lines indicate the threshold value of the CCE norm ($\epsilon^{\rm thres}=0.20$).}
\label{fig:parity}
\end{figure}

We now turn to a detailed examination of local environment around each monomer and to the identification of the crystalline structure (or the lack thereof) at various $T$, mainly in the regime of $T_{cryst}$ and at $T=0$. As described in the methods section, we have implemented the CCE norm and compared against the HCP, FCC, and fivefold structures. The highly discriminating nature of the CCE norm is demonstrated in Figure \ref{fig:parity} where parity plots\cite{cce09} for the HCP and FCC CCE norms are shown for $|\dot{T}|=10^{-6}$ at $T=1.0$, where the system is amorphous, and at $T=0$ where it becomes predominantly ordered. Sites with HCP- (or FCC-) CCE norms with values lower than $\epsilon^{\rm thres}=0.20$ are characterized as HCP-like (or FCC-like). By construction, a monomer with high HCP similarity (i.e. low HCP-CCE norm) possesses low FCC similarity (high value of FCC-CCE norm) and vice versa. Thus, for any system configuration we can reliably identify the local environment around each monomer with respect to HCP, FCC and fivefold symmetries.
The figure illustrates that in the liquid state at $T=1.0$, essentially no monomers have local FCC or HCP order, while at zero temperature, a very high fraction of sites $(\sim65\%$) possess either FCC or HCP order, with comparable probability.
The remaining 35\% of sites have either five-fold or other local structure, as indicated by the region $(\epsilon^{\rm FCC},\epsilon^{\rm HCP}) > (0.2,0.2)$. In parallel, the vacancy of the region defined by $(\epsilon^{\rm FCC},\epsilon^{\rm HCP}) < (0.2,0.2)$, as seen in both panels, highlights the discriminating character of the CCE-norm descriptor.
As we will show below, this noncrystalline portion of samples consists of stack-faulted domain walls and an amorphous ``interphase'' analogous to that found in real semicrystalline polymers.

\begin{figure}[htbp]
\includegraphics[width=3in]{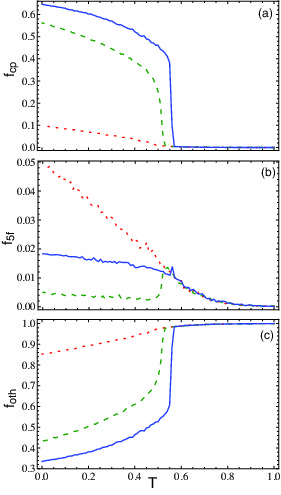}
\caption{Measures of crystalline order vs. $T$ at various quench rates.  Solid blue, dotted green, and dashed red lines show data for $|\dot{T}|=10^{-6}$, $|\dot{T}|=10^{-5}$, and $|\dot{T}|=10^{-4}$, respectively.  Panel (a): fraction of sites $f_{cp}$ with close-packed order, Panel (b): fraction of sites $f_{5f}$ with fivefold local symmetry,  Panel (c): fraction of sites $f_{oth}$ with other local structure.}
\label{fig:metrics}
\end{figure}

Next we present results, based on the CCE analysis, on the evolution of local ordering during cooling.  Figure \ref{fig:metrics} shows the fraction of sites with: (a) close-packed (FCC or HCP) order, (b) fivefold similarity, and (c) neither fivefold local symmetry nor close-packed order, as a function of $T$ for the various quench rates.   In all cases, close-packed ordering grows continuously as $T$ decreases, with the transition at $T=T_{cryst}$ becoming increasingly first-order-like with decreasing $|\dot{T}|$.  As $T$ continues to decrease, the fraction of close-packed sites continues to increase, indicating an effective ``annealing'' process wherein defects are removed.  Throughout this process, the fraction of sites with FCC order is comparable to but exceeds the fraction of sites with HCP order, especially at the slowest $|\dot{T}|$.  This is expected, since while the free energy difference between FCC and HCP phases is very small\cite{bolhuis01}, crystal-growth kinetics favor FCC crystallite formation.\cite{vandeWaal91}

The fraction of sites with close-packed order, $f_{cp}$, increases sharply with decreasing quench rate for $T$ slightly below $T_{cryst}$, and continues to increase as $T$ decreases to zero.
For example, at $T=0$, $f_{cp}$ is only 10\% for $|\dot{T}|=10^{-4}$, but 58\% for $|\dot{T}|=10^{-5}$ and 65\% for $|\dot{T}|=10^{-6}$.
At the intermediate quench rate, the jumps in $U$, $\phi$, and $f_{cp}$ all exhibit a ``delay'' to $T \simeq .52$, indicating that the critical nucleation rate $|\dot{T}|_{crit} \simeq 10^{-5}$.
Quench-rate-dependent differences in crystal structure for $T=0$ will be examined in more detail below.

Fivefold local symmetry is well-known to inhibit crystallization and promote amorphous structure.\cite{frank52,tanaka01,tanaka02,tanaka03,nck01,nck02}  For all quench rates, it is clear that as density $(\phi)$ increases for $T>T_{cryst}$, so does the population of sites with fivefold symmetry. This trend is in perfect agreement with past findings from simulations on  monomeric hard spheres of uniform size\cite{nck02} where fivefold probability in amorphous packings increases as the system becomes denser. The physical trend changes drastically as temperature reaches and drops below $T_{cryst}$.  For the fastest quench rate, fivefold sites continue to grow linearly and the system remains amorphous with only a small fraction of ordered sites. In contrast, the population of fivefold sites drops significantly for $|\dot{T}|=10^{-5}$, and remains nearly constant for $|\dot{T}|=10^{-6}$.  This finding clearly points towards a structural competition between close-packed ordering and fivefold symmetry (e.g.\ twin defects), a physical trend that has also been observed in a wide range of athermal hard-sphere packings.\cite{tanaka01,tanaka02,tanaka03,nck01,nck02}.

Many sites lack either close-packed order or five-fold similarity; panel (c) shows the fraction of such sites, $f_{oth} = 1 - f_{cp} - f_{5f}$.
We note that $f_{oth} \simeq 1$ above $T_{cryst}$, indicating that for the CCE-norm structure-identification procedure described above, $f_{oth}$ is a good discriminant of liquid-like order.
For $|\dot{T}| >  |\dot{T}|_{crit}$, systems retain amorphous structure down to $T=0$, consistent with the $g(r)$ data shown in Fig. \ref{fig:gofr}.
For $|\dot{T}| < |\dot{T}|_{crit}$, $f_{oth}$ shows a first-order-like (downward) jump at $T=T_{cryst}$, and continues to decrease with decreasing $T$ during ``annealing'', but remains significant down to $T=0$.
Our model is therefore well-suited to producing the \textit{semicrystallinity} observed in real polymers.

\begin{figure}[htbp]
\includegraphics[width=3.375in]{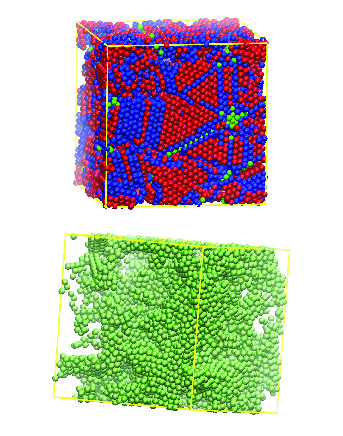}
\caption{Snapshot of system quenched at $|\dot{T}|=10^{-6}$, at $T=0$.  (Top panel) HCP-ordered sites are shown in blue, FCC-ordered sites in red, and fivefold sites in green. (Bottom panel) ``Other'' sites are shown in green. Image created with the VMD software.\cite{vmd01}}
\label{fig:sm6}
\end{figure}

Visualization of systems prepared at various $|\dot{T}|$ provides considerable insight into the semicrystalline morphologies formed by our model.
Figure \ref{fig:sm6} shows the end state $(T=0)$ of the slowest quench.  Grain-like HCP and FCC domains are clearly visible, and the fivefold-symmetric sites often correspond to twin defects - a structure similar to that found in polycrystalline metallic or colloidal systems (see e.g\ Ref.\ \cite{omalley03}) and model packings of monomeric hard spheres.\cite{nck01,nck02}  The ordered structures, as established here, show reduced tendency to layer formation (randomly stacked hexagonal close packing) compared to that found for hard-sphere chains in Refs.\ \cite{karayiannis09,karayiannis10,karayiannis13}, presumably because the larger system sizes employed here reduce the influence of the periodic boundaries or because in the athermal systems a strict tangency condition is applied with respect to bond lengths.
Gaps corresponding to the ``other'' sites are clearly visible in the snapshot.  In the bottom panel, we illustrate these sites for the same system.
Careful visual inspection shows that these regions possess nearly close-packed structure and correspond to stack-faulted domain walls.
Since the crystallite domain size is considerably smaller than our simulation cells, these domain walls form a percolating structure.

\begin{figure}[htbp]
\vspace{-10pt}
\includegraphics[width=3.375in,clip=true]{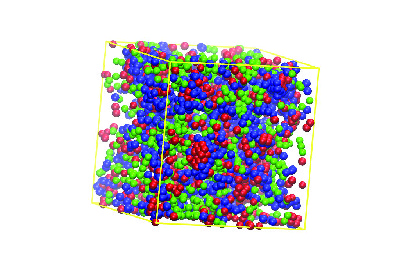}
\vspace{-25pt}
\caption{Snapshot of system quenched at $|\dot{T}|=10^{-4}$, at $T=0$.  HCP-ordered sites are shown in blue, FCC-ordered sites in red, and fivefold sites in green.}
\label{fig:sm4}
\end{figure}

Faster quench rates produce reduced crystalline and greater amorphous order.  Figure \ref{fig:sm4} shows the $(T=0)$ end state of a system quenched at $|\dot{T}|=10^{-4}$.  HCP and FCC crystallites are present, but are far smaller and fewer. Furthermore, the crystallite/grain-boundary structure produced for $|\dot{T}|<|\dot{T}|_{crit}$ is absent for fast quench rates.  In parallel, the number of fivefold sites is much greater than for $|\dot{T}|<|\dot{T}|_{crit}$, and these sites, rather than corresponding to twin defects, are apparently arranged randomly.  Visual inspection of the ``other'' sites for this quench rate shows that they are much less ordered than those for the lowest quench rate, in effect corresponding to an amorphous interphase like that found in real semicrystalline polymers \cite{ward04}.  Thus our model is able to produce large crystallites with domain walls for $|\dot{T}|<|\dot{T}|_{crit}$ and a predominantly amorphous structure with small crystallites for $|\dot{T}|>|\dot{T}|_{crit}$.  For $|\dot{T}|=10^{-5}\simeq|\dot{T}|_{crit}$, results are intermediate between these two limiting cases, with a tendency towards the crystalline state.

\begin{figure}[htbp]
\includegraphics[width=3.375in]{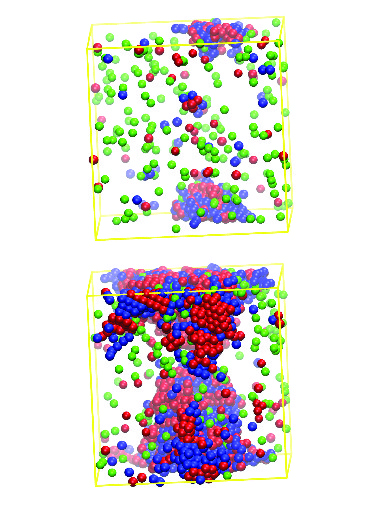}
\caption{Crystal nucleation at $|\dot{T}|=10^-6$.  (Top) Formation of a preliminary nucleus at $T=.562$.  (Bottom) growth of nucleus: $T=.561$.  The systems have $f_{cp}= 0.019$ and $0.14$, respectively.  HCP-ordered sites are shown in blue, FCC-ordered sites in red, and fivefold sites in green. Image created with the VMD software \cite{vmd01}.}
\label{fig:crystnuc}
\end{figure}

Finally, significant information can be gained on crystal nucleation and growth can be obtained by visual examination of systems at $T\simeq T_{cryst}$. Figure \ref{fig:crystnuc} shows a series of snapshots from the $|\dot{T}| = 10^{-6}$ quench. At $T=0.562$ the first trace of crystal aggregates can be seen in the form of a ``baby'' nucleus which consists of similar amounts of HCP and FCC sites.  As the system is still amorphous, the number of sites with fivefold symmetry is comparable to the fraction of sites with either HCP or FCC similarity. However, by $T=0.561$ the number of ordered sites present in the system has greatly increased and the first large crystal seed (critical nucleus) is extant consisting again of roughly equal amounts of HCP- and FCC-like sites. As $T$ continues to drop, this nucleus continues to grow and expand until it fills most of the system as illustrated in Fig.\ \ref{fig:sm6}.

\section{Discussion and Conclusions}
\label{sec:conclude}

We have developed a simple, computationally efficient bead-spring model exhibiting competing crystallization and glass transitions.  Crystallization is promoted by removing the length scale competition present in the Kremer-Grest bead-spring model.\cite{kremer90}  At quench rates faster than the characteristic nucleation rate $|\dot{T}|_{crit}$, systems remain predominantly amorphous with a large number of fivefold-symmetric sites, typifying quenched-disorder vitrification.  At quench rates slower than $|\dot{T}|_{crit}$, the system exhibits a first-order-like phase transition wherein crystal nuclei form in coexistence with an amorphous phase.  Since our model employs flexible chains and possesses a covalent bond length that is equal to the bead diameter, the crystals formed by for slow quench rates exhibit mixed HCP/FCC close-packed order.  While angular interactions and competing length scales produce different crystalline structures in united-atom-model simulations of chemicaly specific polymers, the close-packed order formed by our systems may be viewed as a consequence of the level of coarse-graining; as in the Kremer-Grest model, one bead corresponds to several monomers.
We expect that this model should be useful in studies of how semicrystalline order affects phenomena such as aging, dynamical heterogeneity, and the nonlinear mechanics of solid polymers.

 Our model is comparable to but simpler and computationally ``cheaper'' than a recent bead-spring model of polyvinyl alcohol (PVA)\cite{sommer10,luo13} which has also been used to study generic features of polymer crystallization.
The model of Refs.\ \cite{sommer10,luo13} generates more realistic features of crystallization such as formation of lamellae and chain disentanglement during the crystallization process, but the lamellar equilibrium crystal structure generated by the stiff angular potential employed in the model makes its critical quench rate for crystal formation about 1.5 orders of magnitude (i.e. $\sim10^{-6.5}$) lower than that reported here.
Here we have sacrificed some chemical realism to achieve a minimal model of thermal polymer crystallization.
Future work will develop greater chemical realism and the ability to form lamellar structures, without sacrificing computational efficiency, by integrating generic angular potentials.

\section{Acknowledgments}
\label{sec:results}

We are grateful to Prof. Manuel Laso and Dr. Katerina Foteinopoulou (UPM, Spain) for fruitful discussions on polymer crystallization. NCK acknowledges support by the Spanish Ministry of Economy and Competitiveness (MINECO) through projects "I3" and MAT2010-15482, as well as the computer resources, technical expertise and assistance provided by the Centro de Supercomputacion y Visualizacion de Madrid (CeSViMa).


\begin{thebibliography}{34}
\bibitem[{foo()}]{footdoubpeak}
\bibinfo{note}{Note the slight double peak at $r \sim 1$ in Fig. 2 arises
  because of the slight stiffness mismatch between pair and bond interactions.}
\end{thebibliography}
\end{document}